\documentclass[
 aip,
 amsmath,amssymb,
 reprint,%
]{revtex4-1}

\usepackage[utf8]{inputenc}
\usepackage[T1]{fontenc}

\usepackage{graphicx}
\usepackage{dcolumn}
\usepackage{bm}
\usepackage{mathptmx}
\usepackage[hidelinks]{hyperref}
\usepackage[exponent-product = \cdot,per-mode=fraction]{siunitx}
\DeclareSIUnit\bar{bar}
\DeclareSIUnit\angstrom{\text {Å}}

\usepackage{tikz}
\usetikzlibrary{external}
\usetikzlibrary{shapes, arrows, positioning, fit, calc,decorations.pathreplacing,patterns,math}
\usepackage{mathrsfs}
\usepackage{pgfplots}
\pgfplotsset{filter discard warning=false,compat=newest,/pgf/number format/.cd,1000 sep={}}
\pgfplotsset{every error bar/.style={line width=1pt}}
\usepgfplotslibrary{colormaps}
\usepackage{subcaption,caption}

\tikzstyle{vector} = [draw, -latex']
\tikzstyle{line} = [draw, -]
\tikzstyle{block} = [rectangle,draw, rounded corners,minimum height=2em,minimum width=4em, align = center, shade,left color=lightgray, right color=white, shading angle = -165, font = \scriptsize]
\tikzstyle{arrow} = [draw, -latex', line width = 1pt]
\tikzstyle{marksplot} = [only marks, solid, mark size= 3pt, line width = 0.75pt]
\tikzstyle{linesplot} = [smooth, no markers, line width=1pt]

\def\importfig{true} 

\definecolor{red1}{RGB}{253,212,158}
\definecolor{red2}{RGB}{253,187,132}
\definecolor{red3}{RGB}{252,141,89}
\definecolor{red4}{RGB}{227,74,51}
\definecolor{red5}{RGB}{179,0,0}

\definecolor{green1}{RGB}{204,236,230}
\definecolor{green2}{RGB}{153,216,201}
\definecolor{green3}{RGB}{102,194,164}
\definecolor{green4}{RGB}{44,162,95}
\definecolor{green5}{RGB}{0,109,44}

\definecolor{blue1}{RGB}{208,209,230}
\definecolor{blue2}{RGB}{166,189,219}
\definecolor{blue3}{RGB}{116,169,207}
\definecolor{blue4}{RGB}{43,140,190}
\definecolor{blue5}{RGB}{4,90,141}

\definecolor{gray1}{RGB}{230,230,230}
\definecolor{gray2}{RGB}{180,180,180}
\definecolor{gray3}{RGB}{140,140,140}
\definecolor{gray4}{RGB}{100,100,100}

\definecolor{bigred1}{RGB}{255,247,188}
\definecolor{bigred2}{RGB}{254,227,145}
\definecolor{bigred3}{RGB}{254,196,79}
\definecolor{bigred4}{RGB}{254,153,41}
\definecolor{bigred5}{RGB}{236,112,20}
\definecolor{bigred6}{RGB}{204,76,2}
\definecolor{bigred7}{RGB}{153,52,4}
\definecolor{bigred8}{RGB}{102,37,6}

\definecolor{bigblue1}{RGB}{237,248,177}
\definecolor{bigblue2}{RGB}{199,233,180}
\definecolor{bigblue3}{RGB}{127,205,187}
\definecolor{bigblue4}{RGB}{65,182,196}
\definecolor{bigblue5}{RGB}{29,145,192}
\definecolor{bigblue6}{RGB}{34,94,168}
\definecolor{bigblue7}{RGB}{37,52,148}
\definecolor{bigblue8}{RGB}{8,29,88}

\definecolor{cat1}{RGB}{228,26,28}
\definecolor{cat2}{RGB}{55,126,184}
\definecolor{cat3}{RGB}{77,175,74}
\definecolor{cat4}{RGB}{152,78,163}
\definecolor{cat5}{RGB}{255,127,0}
\definecolor{cat6}{RGB}{255,255,51}
\definecolor{cat7}{RGB}{166,86,40}
\definecolor{cat8}{RGB}{247,129,191}
\definecolor{cat9}{RGB}{153,153,153}

\definecolor{cat0.1}{RGB}{228,26,28}
\definecolor{cat0.2}{RGB}{55,126,184}
\definecolor{cat0.3}{RGB}{77,175,74}
\definecolor{cat0.4}{RGB}{152,78,163}
\definecolor{cat0.5}{RGB}{255,127,0}
\definecolor{cat0.6}{RGB}{255,255,51}
\definecolor{cat1.1}{RGB}{251,180,174}
\definecolor{cat1.2}{RGB}{179,205,227}
\definecolor{cat1.3}{RGB}{204,235,197}
\definecolor{cat1.4}{RGB}{222,203,228}
\definecolor{cat1.5}{RGB}{254,217,166}
\definecolor{cat1.6}{RGB}{255,255,204}

\begin{document}

\title{A Multi-Species Enskog-Vlasov Solver to Determine Evaporation Coefficients of Fluids in High Pressure Environments}

\author{Raphael Tietz}
 \email[Corresponding author: ]{rtietz@irs.uni-stuttgart.de}
\affiliation{
  Institute of Space Systems University of Stuttgart, Pfaffenwaldring 29, 70569 Stuttgart, Germany
}
\author{Rolf Stierle}
  \email{rolf.stierle@itt.uni-stuttgart.de}
\affiliation{
  Institute of Thermodynamics and Thermal Process Engineering, Pfaffenwaldring 9, 70569 Stuttgart, Germany
}
\author{Kim Sophie Ellenberger}
 \email{ellenbergerk@irs.uni-stuttgart.de}
\affiliation{
  Institute of Space Systems University of Stuttgart, Pfaffenwaldring 29, 70569 Stuttgart, Germany
}
\author{Stefanos Fasoulas}
 \email{fasoulas@irs.uni-stuttgart.de}
\affiliation{
  Institute of Space Systems University of Stuttgart, Pfaffenwaldring 29, 70569 Stuttgart, Germany
}
\author{Marcel Pfeiffer}
  \email{mpfeiffer@irs.uni-stuttgart.de}
\affiliation{
  Institute of Space Systems University of Stuttgart, Pfaffenwaldring 29, 70569 Stuttgart, Germany
}

\date{\today, Preprint Submitted to Physics of Fluids.}

\begin{abstract}

This paper introduces a novel multi-species Enskog-Vlasov solver. It is used to determine evaporation coefficients of fluids under high-pressure conditions, a critical factor for efficient fuel mixing in internal combustion engines. The solver handles collisions for different fluid species separately, thereby accurately capturing species-specific interactions essential for realistic evaporation modeling. A new pair correlation function based on the BMCSL equation of state is employed to enhance modeling accuracy, though compliance with Onsager relations remains to be explored.

Validation through various numerical simulations demonstrates the solver's capability. Results from binary fluid relaxation simulations closely match molecular dynamics (MD) data, highlighting accurate collision frequency modeling. Comparative studies against state-of-the-art models verify the solver's precision in predicting liquid-vapor equilibria and evaporation coefficients across diverse pressure scenarios.

Detailed simulations of argon-neon mixtures with realistic particle diameters and masses underscore the significant improvements achieved by employing this multi-species approach over traditional single-species methods. Comparisons with the SAFT-VRQ Mie  and classical density functional theory confirm the solver's reliability in predicting liquid and vapor compositions and detailed density profiles at interfaces over a wide range of pressures and temperatures.

Further analyses illustrate complex dependencies of evaporation and condensation coefficients on temperature and pressure, consistent with existing research findings. Overall, this work advances computational modeling of multi-species evaporation processes in high-pressure environments at different temperatures, providing essential data for improved combustion modeling and broader applications in multi-phase fluid dynamics.
\end{abstract}

\maketitle


\section{Introduction}
\label{sec:Intro}

Human-induced climate change is one of the greatest challenges humanity is facing today \cite{IPCC}. To avoid the worst consequences of climate change, emissions of greenhouse gases such as carbon dioxide must be drastically reduced. One way to achieve this is to replace fossil fuels with electric alternatives, such as replacing internal combustion engines with electric motors. However, for some applications, such as aviation and seafaring, this is hardly possible. Therefore, internal combustion engines in these applications should be as efficient as possible to reduce fuel consumption and carbon emissions.

A major problem with internal combustion engines is their low efficiency, caused in part  by insufficient mixing of liquid fuel with gaseous air. Lamanna \emph{et al.} \cite{Lamanna_2024} proposes a new idea to overcome this problem by heating the fuel droplets above their critical temperature before combustion. This leads to instantaneous disintegration of the fuel droplet due to vanishing surface tension, resulting in better mixing and combustion. They also proposes a model to predict the time until the droplets reach the supercritical state. The problem is, however, that this model relies on evaporation coefficients at pressures and temperatures typical for internal combustion engines, which are not available under these conditions. Therefore, this work aims to provide these coefficients at high pressures and temperatures. In this work, a particle-based Enskog-Vlasov solver is used to determine these needed coefficients. The EV method was chosen because of its similar accuracy and much better efficiency compared to molecular dynamics (MD) for evaporation simulations \cite{frezzotti_mean_2005} especially in performing parameter studies.

The development of numerical solvers for the Enskog equation started in the $1990$s by Montanero \emph{et al.} \cite{montanero_monte_1996,montanero_simulation_1997} and \citet{frezzotti_particle_1997,frezzotti_monte_1999}. In 2005 the latter presented the coupled method for solving the complete EV equation \cite{frezzotti_kinetic_2005,frezzotti_mean_2005}. Since then the method has been used mainly for evaporation simulations \cite{frezzotti_boundary_2011,busuioc_velocity_2020}, but also for non-equilibrium gases \cite{kobayashi_numerical_2012} and polyatomic molecules \cite{busuioc_mean-field_2020}.
The vapor between two liquid phases was also investigated, in isothermal condition \cite{kon_method_2014} as well as in a temperature gradient \cite{kon_kinetic_2017}.
A special kinetic boundary condition for liquid-wall contact has been developed \cite{barbante_kinetic_2012,barbante_kinetic_2015}.
The slip effect of the liquid-vapor interface has also been studied \cite{frezzotti_slip_2012},
as well as round droplets with a 1D spherically symmetric solver \cite{busuioc_weighted_2023,tietz_symmetric_2024}.
A Fokker-Plank method for the EV equation was developed and presented \cite{sadr_treatment_2018,sadr_fokker-planck-poisson_2021}, as well as a BGK model \cite{shan_non-equilibrium_2023}.
The EV equation has also been used to study the evaporation of a binary mixture \cite{frezzotti_mean-field_2018} and evaporation in high pressure environments \cite{ohashi_evaporation_2020} with mechanically identical species. This technique has also been used to study collapsing bubbles in gaseous environments \cite{ohashi_vapor_2023}.

In this work, first the Enskog-Vlasov (EV) equation is introduced in \autoref{sec:EV-eq}. A multi-species Enskog-Vlasov solver has been developed to determine the needed evaporation coefficients and is presented in \autoref{sec:DSMC}. Its implementation can be found in \autoref{sec:PICLas}. Until now, only multi-species field solvers have been developed \cite{frezzotti_mean-field_2018}, but they treat the short-range interaction for all species identically. This paper presents an approach to treat the short-range interactions differently for all species pairs. Due to the spherical symmetry assumption of the EV equation, an argon-neon mixture was studied in this work. The achieved results with this solver are shown in \autoref{sec:Sim}. First, the results are compared with multi-species MD simulations to verify the relaxation times in \autoref{ssec:Reservoir}. The solver is then compared with state-of-the-art results of \citet{ohashi_evaporation_2020} in \autoref{ssec:Compare} with the same species diameter. Then the species diameters and masses are changed to highlight the differences by using species specific data. The solver was also used to predict the liquid and vapor composition of an argon-neon mixture over a temperature range from $\SI{90}{\kelvin}$ to $\SI{142.5}{\kelvin}$ and a pressure range from $\SI{24.5}{\bar}$ to $\SI{98}{bar}$ in \autoref{ssec:SAFT}. The simulation results were afterwards compared with the SAFT-VRQ Mie \citep{aasen_Mie_2019,aasen_mie_2020} and classical density functional theory \cite{hammer_2023}. The evaporation coefficients of this mixture were sampled to determine its temperature and pressure behavior in \autoref{ssec:Evap}.


\section{Enskog-Vlasov equation}
\label{sec:EV-eq}

The Enskog-Vlasov equation \cite{grmela_kinetic_1971} is a fundamental equation in statistical mechanics for dense gases and liquids.
It describes the temporal evolution of the particle distribution function $f(\underline{x},\underline{v},t)$ in position $\underline{x}$, velocity $\underline{v}$, and time $t$. $f$ represents the distribution of particle velocities $\underline{v}$ at a given position $\underline{x}$ and time $t$. The EV equation generalizes the Boltzmann equation by incorporating additional physical effects that account for the intermolecular interactions in dense fluids. This introduces two main modifications to the Boltzmann equation. First, the Vlasov term introduces long-range attractive forces between particles (such as atoms or molecules in the fluid). These forces are essential for modeling stable liquid phases in gaseous environments, as they counteract the increased collision pressure that occurs in dense fluids. Second, the collision integral is replaced by the Enskog collision integral \cite{enskog1922kinetische}, which accounts for the finite size of particles and their increased likelihood of collisions in dense gases. The EV equation is expressed as:

\begin{align}
  \frac{\partial f}{\partial t} + \left<\underline v\cdot\frac{\partial f}{\partial \underline x}\right>+\left<\frac{\underline F}{m}\cdot\frac{\partial f}{\partial \underline v}\right>
  =C_E(f,f),
\end{align}

The left-hand side is a PDE, where $\underline F$ is the force that acts on the particles, leading to an acceleration with the particle mass $m$. The second term is the change of the particle position due to their velocity, and the third one the change of particle velocity due to forces acting on the particles.
On the right-hand side, $C_\mathrm{E}(f,f)$ is the Enskog collision integral. The force $\underline F$ on the particles is a self-consistent mean-field generated by the particles themselves through the spherically symmetric interaction potential $\phi(r)$ with the radius $r$. In the case of the EV equation, the Vlasov integral can be used to determine the mean-field force $\underline F$ onto each particle, which is only evaluated from a minimum distance of one particle diameter $d$, because the short-range interaction is covered by the collision integral. The Vlasov integral reads:
\begin{gather}
  \underline{F} = \int\limits_{\left|\underline{x}_1-\underline x\right|>d} \frac{\mathrm{d}\phi}{\mathrm{d}r} \frac{\underline{x}_1-\underline x}{\left|\underline{x}_1-\underline x\right|} n(\underline{x}_1) \mathrm{d} \underline{x}_1\vphantom{v^{i^{i^{i^{i^i}}}}},
\end{gather}

where $n(\underline{x}_1)$ is the number density at another position $\underline{x}_1$. The derivative of the interaction potential $\frac{\mathrm{d}\phi}{\mathrm{d}r}$ determines the strength of the attraction force at the distance $r$. Therefore, the entire number density field which has a greater distance to $\underline x$ than $d$ acts on the particles. The Sutherland potential is mostly used as an interaction potential, because it mimics the long-range behavior of the Lennard-Jones potential, but it is straightforward to define the particle diameter $d$ by the hard sphere contribution of the Sutherland potential. The Sutherland potential is defined as:
\begin{align}
  \label{eq:Sutherland}
  \phi(r) = \left\{
    \begin{array}{ll}
      +\infty \\
      -\phi_d \left(\frac{r}{d}\right)^{-\gamma}
    \end{array}
    \begin{array}{ll}
      r<d \\
      r\geq d
    \end{array}
  \right.,
\end{align}

where $\phi_d$ is the strength of the potential at distance $d$ which is also the maximum attraction and $\gamma$ is the power factor. In the rest of this paper $\gamma$ is set to $6$ to have the same attraction term as the 12-6 Lennard-Jones potential. Thus, both potentials are consistent with the London dispersion force, which decreases with $r^{-6}\;$ \cite{lennard-jones_cohesion_1931} and is the part of the attractive van der Waals force, which is present for non-dipole species.
This results in similar long-range interactions for both potentials, so evaporation results are also similar \cite{frezzotti_mean_2005}.

The Enskog collision integral $C_\mathrm{E}(f,f)$ includes several effects, which are neglected in the Boltzmann equation. One assumption of the Boltzmann equation is that the mean free path is much greater than the cross-section diameter. Therefore, the distance between the centers of two colliding particles can be neglected. This is not true for high densities, as they occur in liquids, and this displacement of the centers must be considered.
Before proceeding to the description of the Enskog collision integral, it is advisable to take a look at the collision geometry depicted in \autoref{fig:collision}. The Enskog collision integral reads:

\begin{figure}[htb]
  \centering
  
  \ifdefined\importfig
    \includegraphics{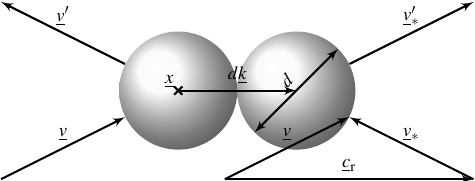}
  \else
    \tikzsetnextfilename{Geo_Collision}
    \input{Geo_Collision}
  \fi

  \caption{Geometry of an Enskog collision. The left particle is at position $\underline x$. Before the collision, it has a velocity of $\underline v$ with is changed during the collision to $\underline{v}'$. The center-point of the second particle is one cross-section diameter $d$ away in the direction of $\underline k$. It has a pre-collision velocity of $\underline{v}_*$ and a post-collision one of $\underline{v}'_*$ The relative velocity $\underline{c}_\text{r}$ is calculated with the pre-collision velocities.}
  \label{fig:collision}
\end{figure}

\begin{align}
  C_\mathrm{E}&(f,f)
  =d^2\int_{\mathcal{S}}\int_{\mathbb{R}^3}R(\big<\underline{c_r}\cdot \underline{k}\big>)\nonumber\\
  &\times\left[
  \begin{array}{ll}
    Y(n(\underline{x}-\frac{d\underline{k}}{2}))f(\underline{x},\underline{v}')f(\underline{x}-d\underline{k},\underline{v}'_*) ...\\
    -Y(n(\underline{x}+\frac{d\underline{k}}{2}))f(\underline{x},\underline{v})f(\underline{x}+d\underline{k},\underline{v}_*)
  \end{array}
  \right]\,\mathrm{d}^3\underline v_*\,\mathrm{d}^2\underline{k},
\end{align}

In the square brackets are the source and sink terms of $f$. In the source term, colliding particles with other velocities will result in the given velocity $\underline v$, while the latter will reduce the number of particles with the given velocity $\underline v$ due to collisions. This means parts of $f$ are only shifted to a different part of the phase space, no particles are deleted or created. The second particle is one cross-section diameter $d$ apart in the direction of $\underline k$. The collision probability is increased by the equilibrium pair correlation function $Y$. It appears in the BBGKY-hierarchy by resolving the two particle distribution functions into a multiplication of two one-particle distribution functions $f_2(\underline{x},\underline{v}_1,\underline{x} + d\underline k,\underline{v}_2) = Y(n(\underline{x}+\frac{d\underline{k}}{2}))f(\underline{x},\underline{v}_1)f(\underline{x}+d\underline{k},\underline{v}_2)$ in dense fluids \cite{karkheck_kinetic_1981}. $Y$ is the contact value of the radial distribution function for a hard sphere fluid. The radial distribution function $g(r)$ returns the relative number density in the vicinity of a given particle compared to the mean number density with respect to the distance to this particle $g(r)=n(r)/n_\text{mean}$. For $r\geq d$, it is a damped oscillating function approaching unity for great distances. For rarefied flows, the radial distribution function and, therefore, its contact value has a value of approximately $1$ and is neglected in the Boltzmann equation.  Due to its strongly damped nature, it is also neglected in the Vlasov integral, where the distances are greater than in the collision process. $R()$ is the ramp function which is defined as $R(a)=\text{max}(a,0)$. Together with the scalar product $\left<\underline{c}_\text{r}\cdot \underline{k}\right>$ it ensures, that only particles which are approaching each other in the direction of $\underline k$ (moving towards each other) are allowed to collide. Otherwise the scalar product would be negative, which leads in combination with $R()$ to a zero and therefore no contribution in the integral. Finally this term is integrated over all incident velocities of the second particle of the collision and the unit sphere spanned by $\underline k$.


\section{Multi-Species Enskog Collision Solver}
\label{sec:DSMC}

\begin{figure}[htb]
  \centering
  
  \ifdefined\importfig
    \includegraphics{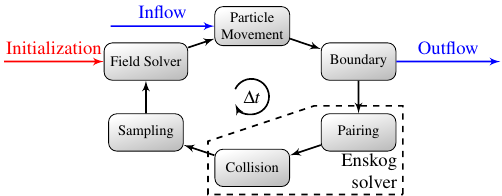}
  \else
    \tikzsetnextfilename{ESMC_Timestep}
    \input{ESMC_Timestep}
  \fi

  \caption{Timestep of the EV-DSMC method. The steps of the Enskog solver are highlighted by the dashed box.}
  \label{fig:ESMC_Scheme}
\end{figure}

Montanero \emph{et al.} \cite{montanero_simulation_1997} and \citet{frezzotti_monte_1999} developed a numerical method to solve the EV equation for a single species of particles. It is based on the DSMC method of Bird \cite{bird_molecular_1994} where the collision solver was adjusted to solve the Enskog collision integral. A flowchart of a EV solver can be seen in \autoref{fig:ESMC_Scheme}, where the \emph{Pairing} and \emph{Collision} steps are conducted by the Enskog solver.
First, the single-species state-of-the-art solver is explained briefly, afterwards the adjustments for the novel multi-species solver are introduced on the basis of the single-species solver.

The single-species solver calculates in the \emph{Pairing} stage for each computational element the necessary number of collision pairs $N_\mathrm{coll}$.  It uses the estimated maximum collision probability $\omega_\text{max}$ and the number of particles in the element $N_\mathrm{part}$ to calculate $N_\mathrm{coll}$:
\begin{equation}
  N_\mathrm{coll} = \frac{1}{2}N_\mathrm{part}\omega_\text{max}
\end{equation}

For each collision pair a random particle in the element and a random unit vector $\underline{k}$ are drawn. At the position $\underline{x}+d\underline{k}$ the second collision particle is searched, where $\underline{x}$ is the position of the first particle. This position can be located in another computational element. Afterwards, in the \emph{Collision} stage, the collision probability $P$ for this pair is calculated. To do this, $Y$ is evaluated with the Carnahan and Starling equation of state (EoS) \cite{carnahan_equation_1969}:
\begin{gather}
  Y(n) = \frac{1}{nb}\left(\frac{p^\text{HS}}{nk_\mathrm{B}T}-1\right) = \frac{1}{2}\frac{2-\eta}{(1-\eta)^3}, \\
  b = \frac{2\pi d^3}{3} ,\;\;\; \eta=\frac{\pi d^3 n}{6}, \nonumber
\end{gather}

where $k_\mathrm{B}$ is the Boltzmann constant. Now $P$ is calculated using the maximum collision probability $\omega_\text{max}$ and the dimensional collision probability $\omega$:
\begin{gather}
  P=\frac{\omega}{\omega_\text{max}},\\
  \omega = R\left(\left<\underline{c}_\text{r}\cdot \underline{k}\right>\right) 4\pi d^2Y\left(\underline{x},\underline{x}+\frac{d\underline{k}}{2}\right)n_{2}\Delta t,\label{eq:Coll_single}
\end{gather}

where $n_2$ is the number density at the position of the second particle and $\Delta t$ is the time step. $P$ is now compared against a random number between $0$ and $1$: $\mathcal{U}(0,1)$. If $P<\mathcal{U}(0,1)$, then no collision will take place, otherwise the velocities of the collision particles are updated to:

\begin{gather}
  \underline{v}' = \underline{v} - \left< \underline{c}_\text{r}\cdot\underline{k} \right>\underline{k}\\
  \underline{v}'_* = \underline{v}_* + \left< \underline{c}_\text{r}\cdot\underline{k} \right>\underline{k}
\end{gather}

This process is conducted in every element for each pair.

A multi-species field solver for the EV equation has already been proposed \cite{frezzotti_mean-field_2018}, but in the Enskog collision solver it handles all species as a single species. In this work, the proposed multi-species Enskog solver is based onto their methods and follows the same workflow as displayed in \autoref{fig:ESMC_Scheme}. This new solver has to consider different masses and diameters of the particles, and therefore, different steps of the original solver have to be altered. In the following the subscripts $i$ and $j$ indicate different species.
One of the main problems of a multi-species solver is that the collision separation has to be known before the second particle is selected, but this separation depends on the species of the second particle. To overcome this issue, the collision process is divided into several passes, one for each species of the second collision particle.

\subsection{Pairing}
In the non-local pairing procedure, the cross-section diameter of the collision pair has to be known, before the second particle can be searched. Therefore, the typical pairing procedure is conducted separately for every species of the second collision particle. Thus, the cross-section diameter is known before selecting the second collision particle, but the second particle has to be selected from the chosen species. The new pairing procedure consists of the following steps:
\begin{enumerate}
  \item[] For every computational element $c$:
  \item Loop over all species. This species defines the species of the $2^\text{nd}$ collision partner $j$:
  \item Calculate the number of collision pairs with species $j$ with
  \begin{equation}
    N_\mathrm{coll} = \frac{1}{2}N_\mathrm{part}\omega_{c,j,\text{max}}
  \end{equation}
  \item[] For every pair:
  \item Draw a random unpaired particle from element $c$, its species is $i$
  \item Draw a random unit vector $\underline k$
  \item Check the path from position of the first particle $\underline x$ to $\underline x + d_{ij} \underline k$ for boundary interactions
  \item Search at position $\underline x + d_{ij} \underline k$ for the nearest neighbor with species $j$
\end{enumerate}

It is interesting to see that $\omega_{c,j,\text{max}}$ can be defined separately for every $j$. This can accelerate simulations if the number densities of the species differ largely and different numbers of collisions with those species are required because of different particle diameters.

\subsection{Collision}
For the collision calculations, the equations are updated for multi-species collisions. In \autoref{eq:Coll_single} only the $n_2$ is altered to the number density of species $j$ in the computational element of the second particle and for $Y$ a multi-species representation has to be used (see \autoref{sec:Pair_Corr}):
\begin{equation}
  \omega = R\left(\left<\underline{c}_\text{r}\cdot \underline{k}\right>\right) 4\pi d_{ij}^2Y\left(\underline{x},\underline{x}+\frac{d_{ij}\underline{k}}{2},i,j\right)n_{c_2,j}\Delta t.
\end{equation}
The collision probability reads:
\begin{equation}
  P = \frac{\omega}{\omega_{c,j,\text{max}}},
\end{equation}
and is compared against $\mathcal{U}(0,1)$ whether a collision occurs or not. The calculation of the post-collision velocities must incorporate the different masses of the species to retain momentum and energy conserving:
\begin{gather}
  \underline{v}' = \underline{v} - 2 \frac{m_i}{m_i+m_j}\left< \underline{c}_\text{r}\cdot\underline{k} \right>\underline{k}\\
  \underline{v}'_* = \underline{v}_* + 2 \frac{m_j}{m_i+m_j}\left< \underline{c}_\text{r}\cdot\underline{k} \right>\underline{k}
\end{gather}

\subsection{Sampling}
The Carnahan and Starling EoS \cite{carnahan_equation_1969} is only valid for a single-species fluid. This EoS was further developed to cover the multi-species case and is known as the BMCSL EoS \cite{boublik_hardsphere_1970,mansoori_equilibrium_1971}.
This EoS can be used to determine the equilibrium pressure of a hard sphere fluid $p^\text{HS}$ and reads:
\begin{gather}
  \frac{p^\text{HS}}{k_\mathrm{B}T}=nZ^\mathrm{HS}=\frac{6}{\pi}\left(\frac{\xi_0}{(1-\xi_3)}+\frac{3\xi_1 \xi_2}{(1-\xi_3)^2}+\frac{(3-\xi_3)\xi_2^3}{(1-\xi_3)^3}\right),\\
  \xi_m = \frac{\pi}{6}\sum_{i=1}^{n_\text{Spec}} n_id_i^m, \label{eq:xi}
\end{gather}

where $T$ is the temperature, and $Z^\mathrm{HS}$ is the compressibility of a hard-sphere fluid, which determines the difference to the ideal gas law. It can be transformed into an analogous equation as in Karkheck \cite{karkheck_kinetic_1981} and Frezzotti \cite{frezzotti_mean_2005}, where the pair contributions of the collision pressure are expanded to account for multi-species. The same can be done with the pair contributions of the attraction field which leads to an equilibrium partial pressure of a species:
\begin{equation}
  p_i = \underbrace{n_ik_\mathrm{B}T\vphantom{\sum_j^{n_\text{Spec}}}}_\text{kinetic} + \underbrace{\frac{2}{3} \pi \sum_j^{n_\text{Spec}}{d_{ij}}^3n_in_jk_\mathrm{B}TY}_\text{collisional} \underbrace{-\frac{2}{3}\pi \frac{\gamma}{\gamma-3}\sum_{j}^{n_\text{Spec}}{d_{ij}}^3\phi_{d,ij}n_in_j}_\text{field},
  \label{eq:press}
\end{equation}
and the complete equilibrium pressure reads:
\begin{equation}
  p = \sum^{n_\text{Spec}}_i p_i.
\end{equation}

For the sampling of non-equilibrium pressure from the particle data, the equations (18) and (20) of Frezzotti \cite{frezzotti_mean_2005} can be used, but it has to be evaluated for every pair of species.

\subsection{Pair Correlation Function}\label{sec:Pair_Corr}
Yau \emph{et al.} proposed a pair correlation function using the BMCSL EoS \cite{yau_pair_1997,boublik_hardsphere_1970,mansoori_equilibrium_1971} which falls back onto the Carnahan and Starling one in the limit of one species:
\begin{align}
  Y(\underline n,d_{i},d_{j}) = \frac{1}{1-\xi_3}&+\frac{3\xi_2}{(1-\xi_3)^2} \left(\frac{d_{i}d_{j}}{d_{i}+d_{j}}\right)\nonumber\\
  &+ \frac{\xi_2^2}{(1-\xi_3)^3}\left(\frac{d_{i}d_{j}}{d_{i}+d_{j}}\right)^2.
\end{align}
In the multi-species case, all $n_i$ and $d_i$ are required to calculate $\xi_m$ as well as the diameters of the two colliding species in order to evaluate the pair correlation function.
The pair correlation function is evaluated at a random position along the line between $\underline x_1$ and $\underline x_1 + d_{ij} \underline k$. As \citet{van_beijeren_modified_1973} noted, a multi-species Enskog solver does not satisfy the Onsager relation if $Y$ is evaluated at the same single point (e.g., always at the midpoint), and proposed an alternative solution to fix this problem. This fix is computationally very demanding, requiring the calculation of an infinite series of Mayer graphs for every collision pair; therefore, it would be difficult to perform such simulations in a reasonable time \cite{benilov_energy_2018}. The compliance of the proposed solver with the Onsager relations has to be investigated in a future work.


\section{Numerical Implementation in PICLas}
\label{sec:PICLas}

The proposed solver is implemented in the PIC-DSMC solver PICLas \cite{fasoulas_combining_2019,tietz_symmetric_2024}. Parts of its architecture can be used for the proposed solver like the particle push, mesh handling, data structures, etc. Besides the Enskog collision solver, several other features are implemented to PICLas to carry out the simulations, like the attraction field solver, sub-cell pairing, relaxation of the collision probability, adaptive pressure boundary conditions, and mixing rules. These are explained in the following.

\subsection{Attraction Field Solver}
The attraction field solver determines the force on the particles. This process includes $3$ steps similar to Particle-in-Cell (PIC) methods \cite{birdsallPlasmaPhysicsComputer1985}: deposition of the particles onto the computational grid, calculation of the force on this grid, and interpolation of the force onto the particle positions.

In the deposition step, the number density is calculated by counting the number of particles inside the element multiplying it by the particle weight $w_\text{k}$ and dividing it by the volume of the element.
The attraction field is similarly handled as in other studies \cite{frezzotti_mean-field_2018,frezzotti_mean_2005}.
For 1D simulations, it is computed by the convolution of the derivative of the interaction potential with the number density field, which results in the force acting on the particles. This convolution can be rewritten as matrix-vector multiplication for every computational element:
\begin{equation}
  \underline{\underline{F}} = \underline{\underline{\underline{\underline{A}}}} \times \underline{\underline{n}} + \underline{\underline{b}}
\end{equation}

The force $\underline{\underline{F}}$ and number densities $\underline{\underline{n}}$ are matrices since they incorporate the 1D position as first dimension and the respective species as second dimension. The Sutherland potential (\autoref{eq:Sutherland}) can be differentiated in $\underline{x}$, after which it is integrated over the two symmetric dimensions. Finally, it is integrated over the $x$-extent of the computational element. The inner repulsive part of the Sutherland potential is excluded and replaced by the potential value of $\phi_d$ since this part is solved by the Enskog collision solver. Due to the discontinuity of the Sutherland potential, the resulting term has three different components: One below of the exclusion region $A_1$ where $r<-d$, one within it $A_2$ where $-d<r<d$, and the last one above of it $A_3$ where $r>d$. Let $x_1 = x_{c,\text{min}} - x_{d,\text{mid}}$ and $x_2 = x_{c,\text{max}} - x_{d,\text{mid}}$
$A_1$ reads as:
\begin{equation}
  A_1(c_2,c_1,i,j) = \left\{
  \begin{array}{ll}
    \frac{2\pi d_{ij}^\gamma}{2-\gamma}(x_1^{2-\gamma}+\tilde{x}_2^{2-\gamma}) & x_1<x_{c_2,\text{mid}}-d_{ij} \\
    0 &\text{otherwise}
  \end{array}
  \right.\hspace{-0.65em},
\end{equation}
where $\tilde{x_2} = \text{min}(x_2,x_{c_2,\text{mid}}-d_{ij})$. $A_2$ reads:
\begin{equation}
  A_2(c_2,c_1,i,j) = \left\{
  \begin{array}{ll}
    \pi (\tilde{x}_1^2-\tilde{x}_2^2) & x_1<x_{d,\text{mid}}+d_{ij}\\& \wedge x_2>x_{c_2,\text{mid}}-d_{ij}\\
    0 &\text{otherwise}
  \end{array}
  \right.\hspace{-0.65em},
\end{equation}
where $\tilde{x_1} = \text{max}(x_1,x_{c_2,\text{mid}}-d_{ij})$ and $\tilde{x_2} = \text{min}(x_2,x_{c_2,\text{mid}}+d_{ij})$. Finally, $A_3$ reads:
\begin{equation}
  A_3(c_2,c_1,i,j) = \left\{
  \begin{array}{ll}
    \frac{2\pi d_{ij}^\gamma}{2-\gamma}(\tilde{x}_1^{2-\gamma}+x_2^{2-\gamma}) & x_2>x_{c_2,\text{mid}}+d_{ij} \\
    0 &\text{otherwise}
  \end{array}
  \right.\hspace{-0.65em},
\end{equation}
where $\tilde{x_1} = \text{max}(x_1,x_{c_2,\text{mid}}+d_{ij})$.
\begin{equation}
  A = \phi_{d,ij}\left(A_1+A_2+A_3\right)
\end{equation}
Evidently, $A$ only depends on the interaction potential and mesh geometry and can therefore be precomputed during the initialization of the simulation code. The same equations can be used to apply boundary conditions. Symmetric and periodic boundaries can be included in $\underline{\underline{\underline{\underline{A}}}}$ where boundaries which are not dependent on the number density field can be included in $\underline{\underline{b}}$.

In the last PIC step, the force is interpolated on the position of the particles.
This is  conducted by a $2^\text{nd}$ order approach. It uses the computational cell containing the particle and its two adjacent ones for a quadratic fit.

\subsection{Sub-cell Pairing}
The density in the vapor phase mainly determines the constant particle weight $\omega_k$ of the simulation. This leads to a high number of particles in the liquid region, due to the density jump in the liquid-vapor interface region.
In the pairing stage, the numeric efficiency of the nearest neighbor search for the second collision particle depends on the total number of particles in the cell, which is an $\mathcal{O}(N^2)$ algorithm. To maintain an efficient code, the computational cells with many particles are subdivided by an 1D Octree algorithm.

For the Enskog DSMC Octree, it is not sufficient to build it
only for the current cell, due to the non-locality of the collision operator.
It has to be built globally beforehand, the pairing and searches have to be conducted on it for all collisions in the computational domain, and it has to be deleted after the whole pairing procedure. In the building process, every cell is checked if the number of particles is higher than $20$. If that is the case, the cell is split into a left and right cell. Each sub-cell is checked again for the particle number criteria and is split again if needed. This procedure is conducted recursively until every sub-cell fulfills the criteria. In the nearest neighbor search, the tree structure is followed through, until the smallest sub-cell containing the position of the second particle ($x_1 + d_{ij} \underline k$) is reached. In this sub-cell the nearest neighbor search to this position is conducted.

\subsection{Maximum Collision Probability Estimator}
For the calculation of the number of collision pairs and the collision probability, the maximum collision probability $\omega_{j,\text{max}}$ has to be estimated. In this work, the estimator is based on the last collision probabilities, which are exponentially smoothed with a margin. In each time step and cell the maximum $\omega_{j}$ is stored. During the update process, $\omega_{j}$ is compared to $\omega_{j,\text{max}}^\text{old}$. If $\omega_{j}$ is larger, then $\omega_{j,\text{max}}^\text{new} = 1.3\omega_{j}$, otherwise it is relaxed with $\omega_{j,\text{max}}^\text{new}= (1-\alpha)\omega_{j,\text{max}}^\text{old} + \alpha 1.3\omega_{j}$; $\alpha = 0.0001$ in this work. The factor of $1.3$ is intended to increase the probability that $\omega_j$ is lower than $\omega_{j,\text{max}}^\text{old}$.

\subsection{Adaptive Pressure Boundary}
To maintain a stable pressure in the computational domain, a new adaptive boundary condition was developed. This adjusts the number density of an inflow boundary \cite{GARCIA2006693} of one species to match a predefined pressure. To achieve this, the number densities of all species in the element next to this boundary are determined. Those number densities are exponentially smoothed. The number density of a predefined species is optimized with \autoref{eq:press} using a Newton algorithm to reach the target pressure. The inflow number density over this boundary is changed to the resulting number density of this species. Particles of this species are removed if they interact with the boundary, for all other species the boundary is treated as an ordinary symmetric boundary.

\begin{figure*}[t]
	\setcounter{subfigure}{0}
	\centering
	\subfloat[Low temperature case, $d_2/d_1=1$]{
  \ifdefined\importfig
    \includegraphics{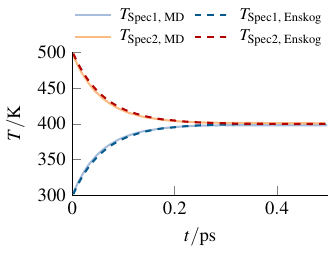}
  \else
    \tikzsetnextfilename{MD_Coll_Model1_cold}
    \input{MD_Coll_Model1_cold}
  \fi
}\hspace{-1em}
    \subfloat[Low temperature case, $d_2/d_1=2$]{
  \ifdefined\importfig
    \includegraphics{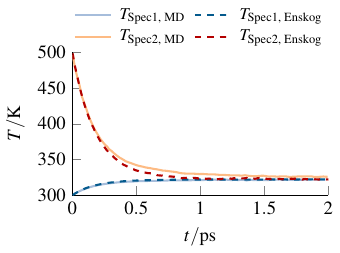}
  \else
    \tikzsetnextfilename{MD_Coll_Model2_cold}
    \input{MD_Coll_Model2_cold}
  \fi
}\hspace{-1em}
    \subfloat[Low temperature case, $d_2/d_1=3$]{
  \ifdefined\importfig
    \includegraphics{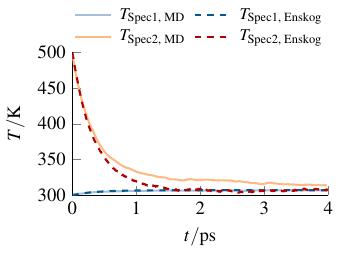}
  \else
    \tikzsetnextfilename{MD_Coll_Model3_cold}
    \input{MD_Coll_Model3_cold}
  \fi
}\\
	\subfloat[High temperature case, $d_2/d_1=1$]{
  \ifdefined\importfig
    \includegraphics{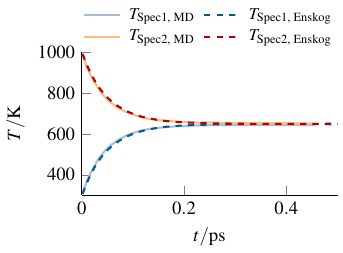}
  \else
    \tikzsetnextfilename{MD_Coll_Model1_hot}
    \input{MD_Coll_Model1_hot}
  \fi
}\hspace{-2em}
    \subfloat[High temperature case, $d_2/d_1=2$]{
  \ifdefined\importfig
    \includegraphics{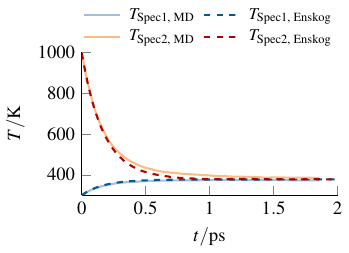}
  \else
    \tikzsetnextfilename{MD_Coll_Model2_hot}
    \input{MD_Coll_Model2_hot}
  \fi
}\hspace{-2em}
    \subfloat[High temperature case, $d_2/d_1=3$]{
  \ifdefined\importfig
    \includegraphics{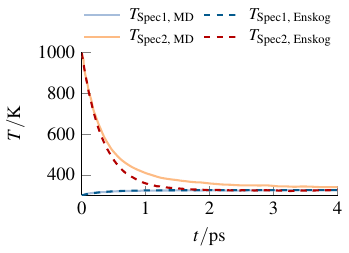}
  \else
    \tikzsetnextfilename{MD_Coll_Model3_hot}
    \input{MD_Coll_Model3_hot}
  \fi
}
	\caption{Binary fluid relaxation in a reservoir compared to MD reference \cite{tanaka_temperature_2020}. The initial temperature for species $1$ is \SI{300}{\kelvin} whereas the temperature for the second species is \SI{500}{\kelvin} for the low temperature case and \SI{1000}{\kelvin} for the high temperature case. The second species is altered in the different cases starting with the same mass and diameter than species $1$, over doubled diameter and $8$ times the mass to $3$ times the diameter and $27$ times the mass of species $1$. All simulations are carried out with a volume fraction of the particles of $0.364$. For the EV simulations the Field solver (Vlasov-Term) was disabled.}
	\label{fig:CollMD_Mod3}
\end{figure*}

\subsection{Mixing rules}
If not stated otherwise, in this work the Lorentz-Berthelot mixing rules are used to calculate the interspecies collision data out of the single species data:
\begin{align}
  d_{ij}&= \frac{d_i+d_j}{2} \\
  \phi_{d,ij}&=\sqrt{\phi_i\phi_j}
\end{align}


\section{Simulations and Discussion}
\label{sec:Sim}

The proposed EV solver is used to perform several types of simulations. The first simulations are intended to validate the solver against results from literature. The single species solver was already successfully verified against literature results \cite{tietz_symmetric_2024}. In the first set of simulations, the relaxation of a binary fluid in a reservoir is simulated and compared with MD results. Then, the solver is tested against state-of-the-art solvers to predict the composition of a binary fluid and its vapor and evaporation in a high-pressure environment where the species diameter and mass are the same for all species. In the second case, the properties of the second species are changed to highlight the difference when using a multi-species Enskog solver. The solver is then used to simulate the composition of an argon fluid in a neon environment at different pressure levels and compared to the SAFT-VRQ Mie EoS (Statistical Associating Fluid Theory for a Mie potential for Quantum fluids with a Variable Range) \cite{aasen_Mie_2019,aasen_mie_2020}. Finally, the evaporation and condensation coefficients of these simulations are sampled.

\subsection{Reservoir Relaxation}
\label{ssec:Reservoir}

Tanaka \emph{et al.}\cite{tanaka_temperature_2020} performed MD simulations to determine how fast an excited molecule produced by chemical reactions must equilibrate in biological cells. They simulated a binary mixture of hard spheres. The first species represents water, while the second is modified by a factor of $2$ and $3$ in diameter and $8$ and $27$ in mass, respectively. The total packing fraction is about \SI{40}{\percent} and both species occupy the same volume fraction. The initial temperature for the first species was always \SI{300}{\kelvin}, while the temperature of the second species is either \SI{500}{\kelvin} or \SI{1000}{\kelvin}. This results in a total of six simulations. The Vlasov-term was disabled because in this test case are no attractive fields present.

These simulations are used to test the relaxation rate of the new solver in a binary fluid and are shown in \autoref{fig:CollMD_Mod3}. The translational temperatures of the new Enskog solver are close to the MD simulations for both species in the relaxation process. At equilibrium the MD simulations (especially for larger species) do not converge to equilibrium. Tanaka \emph{et al.}\cite{tanaka_temperature_2020} state that they face some convergence problems due to the low number of particles especially in the cases with a large size ratio of the species. Since the temperatures in the relaxation phase are similar, the collision frequency and the collision treatment could replicate the processes well.

\subsection{Comparison with the State of the Art}
\label{ssec:Compare}

As mentioned in the introduction, \citet{frezzotti_mean-field_2018} and \citet{ohashi_evaporation_2020} have already performed intermediate multi-species EV simulations. They implemented a multi-species field solver, but treated all particles as single species in the Enskog collision solver. The first publication \cite{frezzotti_mean-field_2018} deals with the evaporation of a binary fluid in a vacuum, while the second \cite{ohashi_evaporation_2020} deals with evaporation in a high-pressure environment, where the second species exerts the pressure on the liquid bulk. The second scenario is more in line with the scope of this work, and therefore the solver will be tested against their results.

The simulations were performed at a rather low temperature of $\num{0.633} T_\text{c,Ar}$. The critical temperature of neon is set to $T_\text{c,Ne} = \SI{11.38}{\kelvin}$, which is about a factor of $4$ lower than in reality, to avoid the perturbation of the liquid bulk by too much dissolved neon. This is indicated by placing this special `neon' in quotes.
The strengths of attraction used in these sets of simulations are $\frac{\phi_{d,\text{Ar}}}{k_\text{B}} = \SI{199.72}{\kelvin}$, $\frac{\phi_{d,\text{Ne}}}{k_\text{B}} = \SI{15.073}{\kelvin}$, and $\frac{\phi_{d,\text{Ar,Ne}}}{k_\text{B}} = \frac{\phi_{d,\text{Ne,Ar}}}{k_\text{B}} = \SI{54.866}{\kelvin}$. The spatial resolution is $0.1 d_\text{Ar}$ and the particle weight is $4.3\times10^{15}$. The size of the computational domain in the x-direction is set to $30 d_{Ar}$, while the $y$- and $z$-directions have a size of unity. The liquid domain is initialized with \num{13440} pure argon particles, while the gaseous domain is initialized with varying amounts of neon particles: \SI{5}{\percent} of the argon particles, \SI{13}{\percent}, \SI{25}{\percent}, \SI{50}{\percent}, \SI{100}{\percent}, \SI{125}{\percent}, \SI{150}{\percent}, and \SI{200}{\percent} of them.

\begin{figure}[htb]
  \centering
  \vspace*{-0.5em}
  
  \ifdefined\importfig
    \includegraphics{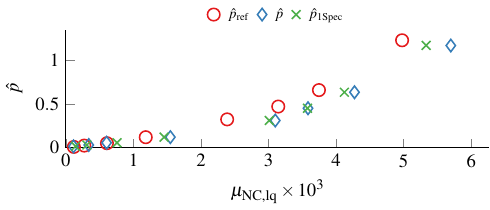}
  \else
    \tikzsetnextfilename{Henris_Law}
    \input{Henris_Law}
  \fi

  \vspace*{-0.5em}
  \caption{Dissolved `neon' content at respective dimensionless pressure $\hat{p} = p d^3/k_\text{B} T_\text{c,Ar}$ compared to the reference \cite{ohashi_evaporation_2020}. The crosses are the results of the method of the reference included into the PICLas framework, where the collsisions are treated if there is only a single species.}
  \label{fig:comp_ohashi_comp}
  \vspace*{-0.5em}
\end{figure}

The results are shown in \autoref{fig:comp_ohashi_comp}. The pressure agrees well with the reference, while the dissolved `neon' content has a deviation of about \SI{20}{\percent}, but follows the same trend. The method of the reference \cite{ohashi_evaporation_2020} was included in the PICLas framework and shows almost the same results as the full multi-species simulations.

\begin{figure}[!htb]
  \centering
  
  \ifdefined\importfig
    \includegraphics{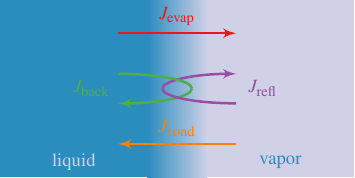}
  \else
    \tikzsetnextfilename{Evap_Fluxes}
    \input{Evap_Fluxes}
  \fi

  \caption{Definition of particle fluxes across the liquid-vapor interface. Particles entering the interface region from the liquid bulk can either evaporate $J_\text{evap}$ or fall back into the liquid $J_\text{back}$. The same is true for particles coming from the vapor region, they can either condensate $J_\text{cond}$ or be reflected from the liquid $J_\text{refl}$.}
  \label{fig:evap_fluxes}
\end{figure}

In evaporation simulations the evaporation and/or condensation coefficient is usually determined. They can be used to determine the evaporation flux \cite{frezzotti_mean_2005,frezzotti_mean-field_2018} or to formulate kinetic boundary conditions \cite{kobayashi_kinetic_2017}. The evaporation coefficient $\sigma_\text{e}$ and its condensation pendant $\sigma_\text{c}$ can be expressed by particle fluxes crossing the interface (see \autoref{fig:evap_fluxes}):
\begin{align}
  \sigma_{\text{e},i}=\frac{J_{\text{evap},i}}{J_{\text{evap},i}+J_{\text{refl},i}} = \frac{J_{\text{evap},i}}{J_{\text{out},i}}, \\ \sigma_{\text{c},i}=\frac{J_{\text{cond},i}}{J_{\text{cond},i}+J_{\text{refl},i}} = \frac{J_{\text{cond},i}}{J_{\text{in},i}},
\end{align}

where the subscript $i$ indicates the species.

The different fluxes can be sampled directly in the simulation by tagging particles as they cross the liquid or vapor boundary of the interface region. Unfortunately, $\sigma_\text{e}$ and $\sigma_\text{c}$ depend on the defined position of this boundary.
To circumvent this problem, Ishiyama \emph{et al.} \cite{ishiyama_molecular_2004} developed a sampling method. He proposed to sample the particle fluxes directly in the simulation and defined the liquid and vapor sampling plane as follows:
\begin{gather}
  n_\text{fit}(x)=\frac{n_{\text{vap}}+n_\text{liq}}{2}+\frac{n_{\text{vap}}-n_\text{liq}}{2}\text{tanh} \left( \frac{x-X_\text{m}}{0.455\delta}\right), \label{eq:Int_fit}\\
  x_\text{liq} = X_\text{m} - \delta, \;\;\; x_\text{vap} = X_\text{m} + 3 \delta,
\end{gather}
where $X_m$ is the center of the interface and $\delta$ is its 10-90 thickness. Both are determined by fitting the upper expression to the number density profile along the interface region. $x_\text{liq}$ and $x_\text{vap}$ define the sampling plane at the interface in the respective phases. The vapor boundary must be further away to cover the entire potential well of the liquid. Placing it closer would cause particles falling back into the liquid ($J_\text{back}$) to be counted as evaporating and then as condensing, and would distort $\sigma_\text{e}$ and $\sigma_\text{c}$.

\begin{figure}[b]
  \centering
  \subfloat[Argon]{
  \ifdefined\importfig
    \includegraphics{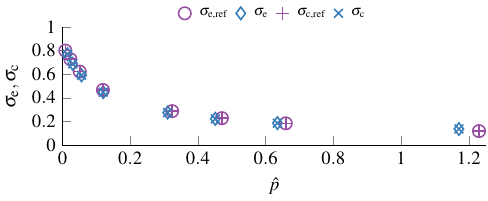}
  \else
    \tikzsetnextfilename{Evap_Coeffs_Ar}
    \input{Evap_Coeffs_Ar}
  \fi
}\\
  \subfloat[`Neon']{
  \ifdefined\importfig
    \includegraphics{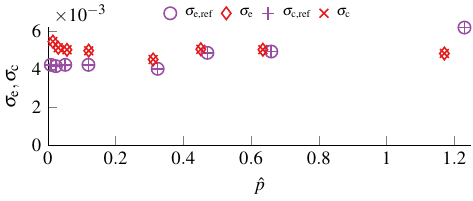}
  \else
    \tikzsetnextfilename{Evap_Coeffs_Ne}
    \input{Evap_Coeffs_Ne}
  \fi
\label{fig:ohashi_neon_evap}}
  \caption{Evaporation and condensation coefficients compared to \citet{ohashi_evaporation_2020}.}
  \label{fig:comp_ohashi}
\end{figure}

For the simulations, $\sigma_\text{e}$ and $\sigma_\text{c}$ were evaluated and compared with the reference as shown in \autoref{fig:comp_ohashi}. The authors would like to note that `neon' cannot condense or evaporate; rather, it dissolves or degases from the liquid. However, since this process can be sampled in the same way as argon's evaporation and condensation, it is also referred to as  $\sigma_\text{e}$ and $\sigma_\text{c}$ as well in this paper. The results are in good agreement with the reference, with slight deviations for argon. The evaporation coefficient of `neon' has higher deviations. It is higher than the reference at low pressures and converges to the reference at higher pressures.

\subsection{Extended Simulations with Different Particle Diameters}

\begin{figure}[b]
  \centering
  \subfloat[Argon liquid and `neon' vapor density.]{
  \ifdefined\importfig
    \includegraphics{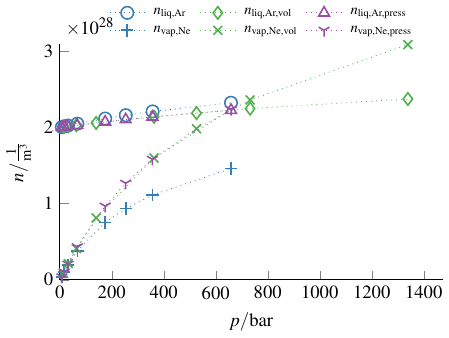}
  \else
    \tikzsetnextfilename{NumDens_extended}
    \input{NumDens_extended}
  \fi
}\\
  \subfloat[`Neon' liquid and argon vapor density.]{
  \ifdefined\importfig
    \includegraphics{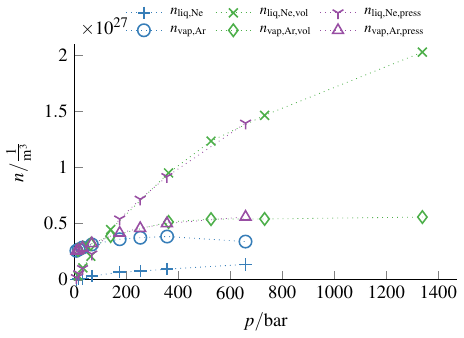}
  \else
    \tikzsetnextfilename{NumDens_extended_trace}
    \input{NumDens_extended_trace}
  \fi
}
  \caption{Liquid and vapor number densities for the extended simulations with different particle diameters and masses. The blue marks are the results of the previous simulations. The other data points are from simulations with different masses and diameters of argon and `neon'. The subscript `vol' adjusts the inserted number density of `neon' to match the particle volume of the previous simulations. The subscript `press' corresponds to the pressure of the previous simulations. The dotted lines are eye-guides to connect data of the same type.}
  \label{fig:ohashi_extended_n}
\end{figure}

The next set of simulations is conducted with the real mass and diameter of `neon' to highlight the differences between the single-species and multi-species approaches. Two additional sets of simulations are performed, one with the same volume fraction of `neon' and one with the same pressure as in the previous simulations. For the latter set, the adaptive pressure boundary was used. The mass and diameter of argon was set to \SI{6.634e-26}{\kilogram} and \SI{3.332}{\angstrom} and those of `neon' to \SI{3.351e-26}{\kilogram} and \SI{2.566}{\angstrom}, respectively.

\begin{figure}[b]
  \centering
  \subfloat[Argon]{
  \ifdefined\importfig
    \includegraphics{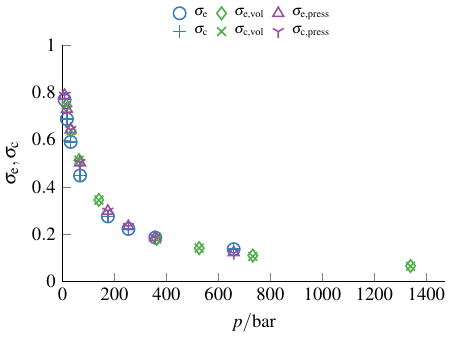}
  \else
    \tikzsetnextfilename{Variable_Evap_Coeffs_Ar}
    \input{Variable_Evap_Coeffs_Ar}
  \fi
}\\
  \subfloat[`Neon']{
  \ifdefined\importfig
    \includegraphics{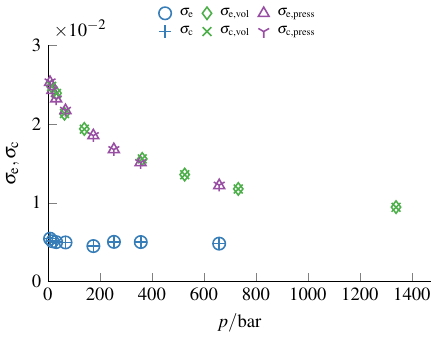}
  \else
    \tikzsetnextfilename{Variable_Evap_Coeffs_Ne}
    \input{Variable_Evap_Coeffs_Ne}
  \fi
}
  \caption{Evaporation and condensation coefficients for the extended simulations with variable particle diameters and masses. $\sigma_\text{e}$ and $\sigma_\text{c}$ are the dimensionalized results of \autoref{fig:comp_ohashi}. The other data points are from simulations with different masses and diameters of argon and `neon'. The subscript `vol' adjusts the inserted number density of `neon' to match the particle volume of the previous simulations. The subscript `press' corresponds to the pressure of the previous simulations.}
  \label{fig:ohashi_extended_sigma}
  \vspace{1.3em}
\end{figure}

The densities in the liquid and vapor phases in the simulations are shown in \autoref{fig:ohashi_extended_n}. It can be seen that the number density of neon in the gas phase is higher than in the case with equal diameters and masses. This is due to the smaller diameter of 'neon', therefore more neon atoms are needed to create the same pressure as in the first case. It is also interesting that much more neon is dissolved in the liquid phase, probably because the collision pressure is lower due to the smaller diameter, and therefore the `neon' can be more easily dissolved in the liquid. There is also more argon in the gaseous phase at higher pressures.
The evaporation and condensation coefficients are shown in \autoref{fig:ohashi_extended_sigma}.
In (a) the results for argon are shown. The difference between the single-species collision solver and the multi-species solver is minimal. In (b) the results for neon are shown, where the differences between the single species collision solver and the multi species solver are significant. Probably due to the smaller cross-sectional diameter of `neon' in the multi-species simulations, `neon' can evaporate more easily, increasing the evaporation coefficient, which now follows a similar trend to that of argon, but at a much lower level. In particular, the evaporation coefficient for `neon' is now not constant but high for low pressures and follows an exponential decrease for higher pressures.
These results highlight the importance of the use of a multi-species collision solver over a single-species solver for such evaporation simulations.

\subsection{Composition of Argon-Neon Mixtures}
\label{ssec:SAFT}

To test the full multi-species capabilities of the solver, the multi-species equilibrium results are compared to the the statistical associating fluid theory for Mie potentials of variable range corrected for quantum effects (SAFT-VRQ-Mie \citep{aasen_Mie_2019,aasen_mie_2020}). SAFT-VRQ Mie approximates the thermodynamic behavior of spherical atoms or molecules, including quantum effects, by a third-order Barker--Henderson perturbation theory for Mie fluids that includes Feynman--Hibbs quantum corrections. At sufficiently high temperatures, it simplifies to SAFT-VR Mie \cite{Lafitte_fluid_2013} where no quantum correction is required. The parameters used and the average absolute deviations of some thermodynamic properties are listed in \autoref{tab:saftvrqmieparameter}. The pure component parameters were taken from the literature \cite{dufal_2015,aasen_Mie_2019} and the binary interaction parameters were adjusted to experimental vapor-liquid equilibrium data.

To predict density profiles at vapor-liquid interfaces, SAFT-VRQ Mie can be extended to a Helmholtz energy functional \cite{hammer_2023} suitable for use within classical density functional theory (DFT).
Classical DFT allows to obtain equilibrium density profiles (as opposed to electron density profiles from quantum DFT) by means of a variational principle. Since no interfacial-specific parameters are used, the results can be regarded as purely predictive.

\begin{table}[htb]
  \caption{Component parameters for SAFT-VRQ Mie where \(m\) is the chain length (argon and neon are spherical), size \(\sigma\) and energy parameter \(\varepsilon\), repulsive and attractive exponents of the Mie potential, \(\lambda_\mathrm{r}\) and \(\lambda_\mathrm{a}\), respectively, and the Feynman--Hibbs (FH) order. The used binary interaction parameters \cite{aasen_mie_2020} are \(k_{ij}\) and \(l_{ij}\).}
  \label{tab:saftvrqmieparameter}
  \begin{tabular}{c|ccccccc}
               & \(m\)   & \(\sigma\)/\si{\angstrom} & \(\frac{\varepsilon}{k_\mathrm{B}}\)/\si{\kelvin} & \(\lambda_\mathrm{r}\) & \(\lambda_\mathrm{a}\) & FH-order & Ref.  \\\hline
    argon      & 1       & \num{3.4040}              & \num{117.84}                                      & \num{12.085}           & \num{6.0}              & 0        & [\!\citenum{dufal_2015}]  \\
    neon       & 1       & \num{2.7778}              & \num{37.501}                                      & \num{13.000}           & \num{6.0}              & 1        & [\!\citenum{aasen_Mie_2019}] \\\hline\hline
    \(k_{ij}\) & \multicolumn{7}{l}{\num{-0.04456543}}              \\
    \(l_{ij}\) & \multicolumn{7}{l}{\num{-0.04481226}}              \\
  \end{tabular}
\end{table}

\begin{table}[htb]
  \caption{The average absolute deviation (AAD\%) of saturation pressure \(p^\mathrm{sat}\), saturated liquid density \(\rho^\mathrm{liq,sat}\), and critical temperature \(T_\mathrm{c}\) of SAFT-VRQ Mie for argon and neon as reported by \citet{dufal_2015}, and \citet{aasen_Mie_2019}.}
  \label{tab:saftvrqmieaad}
  \begin{tabular}{c|ccc|c}
               & \multicolumn{3}{c}{AAD\%} & \\\hline
               & \(p^\mathrm{sat}\) & \(\rho^\mathrm{liq,sat}\) & \(T_\mathrm{c}\) & Ref.                            \\\hline
    argon      & \num{0.21}         & \num{0.66}                & \num{2.19}       & [\!\citenum{dufal_2015}]        \\
    neon       & \num{0.49}         & \num{0.40}                & \num{1.20}       & [\!\citenum{aasen_Mie_2019}]    \\\hline\hline
  \end{tabular}
\end{table}

\begin{figure}[htb]
  \centering
  \subfloat[Fitting argon diameter to liquid density of SAFT-VR Mie \cite{Lafitte_fluid_2013}]{
  \ifdefined\importfig
    \includegraphics{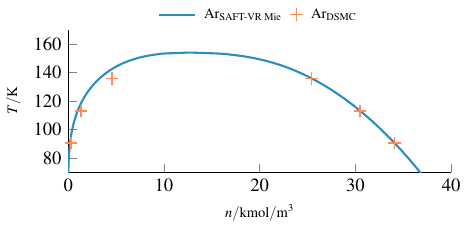}
  \else
    \tikzsetnextfilename{d_vap}
    \input{d_vap}
  \fi
\label{fig:argon_fit}}\\
  \subfloat[Fitting neon diameter to pressure data of NIST \cite{standards_nist_2000}]{
  \ifdefined\importfig
    \includegraphics{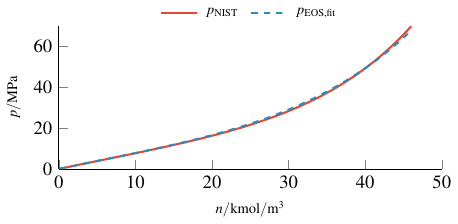}
  \else
    \tikzsetnextfilename{d_nc}
    \input{d_nc}
  \fi
}
  \caption{Determining particle diameters\label{fig:neon_fit}}
  \label{fig:fit_quality}
\end{figure}

For this comparison, the particle diameters of argon and neon have to be determined for the Enskog solver. For argon, the liquid densities of pure argon at several temperatures are fitted to SAFT-VRQ Mie. The quality of this fit is shown in \autoref{fig:argon_fit}. Neon is in a supercritical state in all simulations. Therefore, the pressure of gaseous neon is fitted to the NIST \cite{standards_nist_2000} data by the least squares fit of \autoref{eq:press}. The results are shown in \autoref{fig:neon_fit}.

Four different sets of simulations were carried out. Each set has the same pressure, but with different temperatures. The pressures are \SIlist[list-units=single]{24.5;49;73.5;98}{\bar}, where the last is about $2$ times the critical pressure of argon. The adaptive pressure boundary is used to keep the chosen pressure in the simulation domain. The temperatures are \SIlist[list-units=single]{90;97.5;105;112.5;120;127.5;135;142.5}{\kelvin}. The argon diameters are listed in \autoref{tab:part_dia}, while the neon diameter is kept constant at \SI{2.566}{\angstrom} due to the small temperature effect in this temperature range.

\begin{table}[htb]
  \caption{Particle diameters of argon at the different temperatures.}
  \label{tab:part_dia}
  \begin{tabular}{c|cccccccc}
    $T/\si{\kelvin}$              & $90$    & $97.5$  & $105$   & $112.5$ & $120$   & $127.5$ & $135$   & $142.5$ \\\hline
    $d_\text{Ar}/\si{\angstrom}$ & $3.364$ & $3.321$ & $3.280$ & $3.242$ & $3.207$ & $3.174$ & $3.145$ & $3.117$
  \end{tabular}
\end{table}

The spatial resolution is set to \SI{0.2889}{\angstrom} and the time step to \SI{0.833}{\femto\second}. The simulation domain is initialized with argon particles with a particle density of \SI{1.5e28}{\per\cubic\meter} on the left \SI{60}{\angstrom} side of the computational domain. The rest is filled with neon particles with the number density of the respective pressure and temperature.

\begin{figure*}
  \centering
  \subfloat[\SI{24.5}{\bar} and \SI{90}{\kelvin}   ]{
  \ifdefined\importfig
    \includegraphics{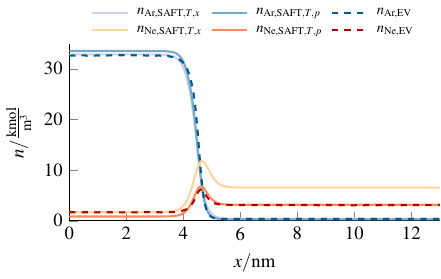}
  \else
    \tikzsetnextfilename{DensProf_24.5_090}
    \input{DensProf_24.5_090}
  \fi
}
  \subfloat[\SI{24.5}{\bar} and \SI{127.5}{\kelvin}]{
  \ifdefined\importfig
    \includegraphics{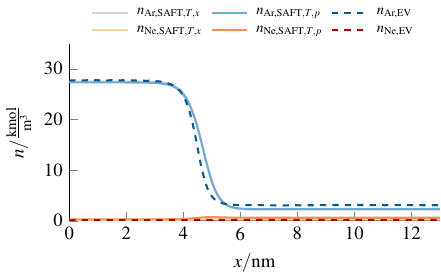}
  \else
    \tikzsetnextfilename{DensProf_24.5_127.5}
    \input{DensProf_24.5_127.5}
  \fi
}\\
  \subfloat[\SI{49}{\bar} and \SI{90}{\kelvin}   ]{
  \ifdefined\importfig
    \includegraphics{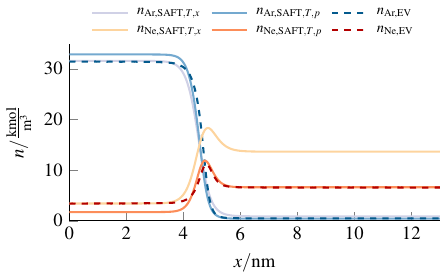}
  \else
    \tikzsetnextfilename{DensProf_49_090}
    \input{DensProf_49_090}
  \fi
}
  \subfloat[\SI{49}{\bar} and \SI{127.5}{\kelvin}]{
  \ifdefined\importfig
    \includegraphics{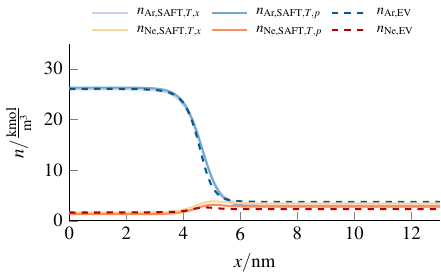}
  \else
    \tikzsetnextfilename{DensProf_49_127.5}
    \input{DensProf_49_127.5}
  \fi
}\\
  \subfloat[\SI{73.5}{\bar} and \SI{90}{\kelvin}   ]{
  \ifdefined\importfig
    \includegraphics{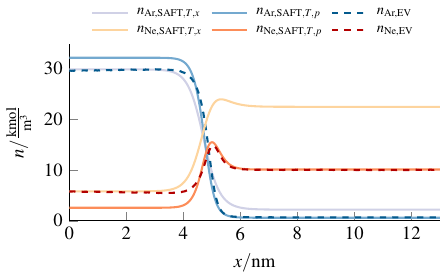}
  \else
    \tikzsetnextfilename{DensProf_73.5_090}
    \input{DensProf_73.5_090}
  \fi
}
  \subfloat[\SI{73.5}{\bar} and \SI{127.5}{\kelvin}]{
  \ifdefined\importfig
    \includegraphics{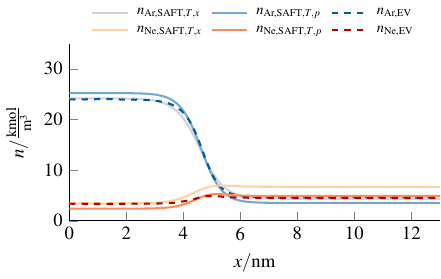}
  \else
    \tikzsetnextfilename{DensProf_73.5_127.5}
    \input{DensProf_73.5_127.5}
  \fi
}\\
  \subfloat[\SI{98}{\bar} and \SI{90}{\kelvin}   ]{
  \ifdefined\importfig
    \includegraphics{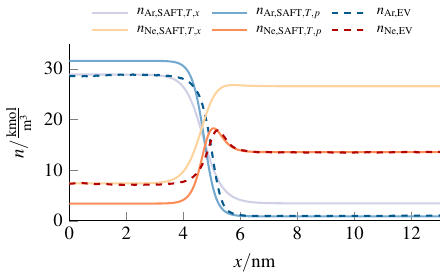}
  \else
    \tikzsetnextfilename{DensProf_98_090}
    \input{DensProf_98_090}
  \fi
}
  \subfloat[\SI{98}{\bar} and \SI{127.5}{\kelvin}]{
  \ifdefined\importfig
    \includegraphics{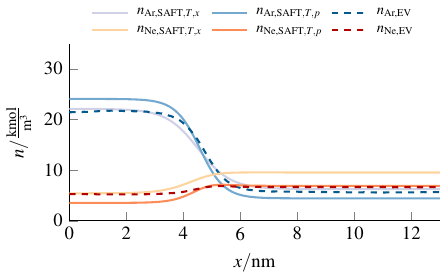}
  \else
    \tikzsetnextfilename{DensProf_98_127.5}
    \input{DensProf_98_127.5}
  \fi
}
  \caption{Density profiles of the argon-neon mixtures compared to two different DFT results. The '\(T, x\)' case's input is the composition of the liquid phase of the EV case, whereas, the '\(T, p\)' case's input is the temperature and pressure of the EV case. The blue lines are the argon profiles and the red ones the profiles of neon.}
  \label{fig:saft_Profiles}
\end{figure*}

\begin{figure*}
  \centering
  \subfloat[\SI{24.5}{\bar}]{
  \ifdefined\importfig
    \includegraphics{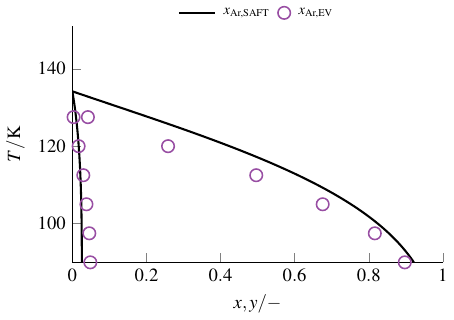}
  \else
    \tikzsetnextfilename{p1}
    \input{p1}
  \fi
}
  \subfloat[\SI{49}{\bar}]{
  \ifdefined\importfig
    \includegraphics{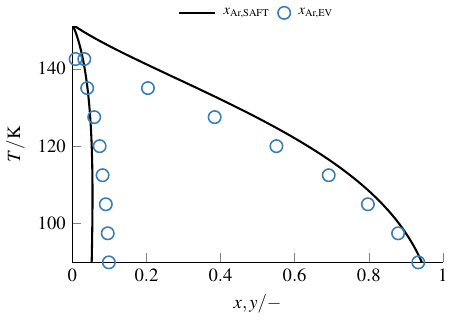}
  \else
    \tikzsetnextfilename{p2}
    \input{p2}
  \fi
}\\
  \subfloat[\SI{73.5}{\bar}]{
  \ifdefined\importfig
    \includegraphics{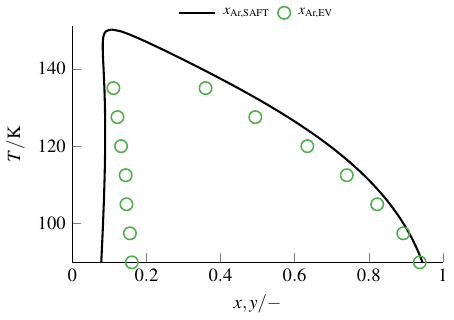}
  \else
    \tikzsetnextfilename{p3}
    \input{p3}
  \fi
}
  \subfloat[\SI{98}{\bar}]{
  \ifdefined\importfig
    \includegraphics{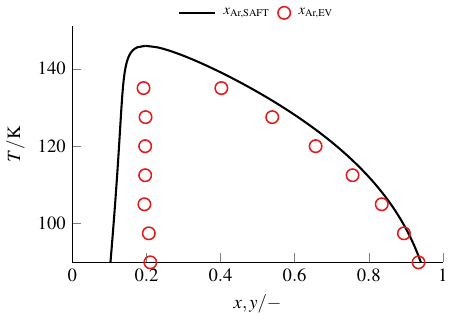}
  \else
    \tikzsetnextfilename{p4}
    \input{p4}
  \fi
}
  \caption{Composition compared to SAFT-VRQ Mie. The solid black lines are the results of SAFT-VRQ Mie, whereas the circles are those from the Vlasov-Enskog solver. The left branch is the liquid state and the right one is the liquid composition, respectively.}
  \label{fig:saft}
\end{figure*}

The density profiles of selected simulations are shown in \autoref{fig:saft_Profiles} and compared to density profiles from classical DFT obtained using FeO\textsubscript{s} \cite{rehner_2023}. The blue argon profiles show a high number density in the liquid and a low one in the vapor region. In the interface layer in between, there is a steep gradient connecting both levels. This behavior can be seen in many previous studies \cite{frezzotti_kinetic_2005,busuioc_velocity_2020}. The neon profiles are displayed in red. Depending on the case, the profile does not simply connect the liquid and gaseous state, but sometimes overshoots them. It is more prone for the cold cases than for the hot ones and can also be seen in previous MD simulations \cite{baidakov_molecular-dynamics_2008,kobayashi_kinetic_2017,nitzke_2023}. If the classical DFT uses the temperature and liquid composition (\(T,x\)-case) of the EV case as input, the liquid phase is in good agreement with the EV solver; if the temperature and pressure (\(T,p\)-case) as the EV solver are used, the vapor phase is in good agreement between the two approaches. This partial match is due to inconsistency between SAFT-VRQ Mie EoS and the mean-field approach of the EV field solver as well as the different interaction potentials. The DFT \(T,p\)-case and EV simulations predict also a similar agglomeration of neon, interface thickness, and gradients in the interface.

The composition of the liquid and gas phase of all simulations are shown in \autoref{fig:saft} and compared with SAFT-VRQ Mie. In the \SI{24.5}{\bar} case, simulations beyond \SI{127.5}{\kelvin} were not possible, because the saturation vapor pressure was higher than the given pressure. In the \SI{73.5}{\bar} and \SI{98}{\bar} case, the liquid bulk vanished completely at a temperature of \SI{142.5}{\kelvin}, the authors assume this happens because the critical temperature of the mixture was exceeded.

The EV and SAFT-VRQ Mie results follow the same trend. In the left branch, the liquid composition, the mole fraction of neon decreases with increasing temperature. At higher pressures, this behavior changes to the opposite in the highest pressure case. The same behavior can be seen for the EV simulations, but weaker. In the highest pressure case, the mole fraction of neon is almost independent of temperature. At low temperatures, the concentration of neon in the liquid phase is almost twice as high as for SAFT-VRQ Mie. The deviation shrinks with increasing temperature, and for the two low pressure cases, the EV results predict a lower dissolved mole fraction than the SAFT results. The dissolved mole fraction increases almost proportionally with pressure.

For the vapor branch in \autoref{fig:saft}, which is on the right side in the plots, the behavior of the difference between the EV and SAFT results is reversed. The lowest deviation is for the lowest temperature case and increases with temperature. In all cases, the SAFT predicts a higher mole fraction of neon in the gaseous region. The vapor composition moves towards the liquid composition with higher temperature, weakly at low temperatures and approaching faster and faster. Most of the difference is probably due to the use of a different interaction potential in the two methods and the mean-field simplification of the EV solver.

\begin{figure}[htb]
  \centering
  
  \ifdefined\importfig
    \includegraphics{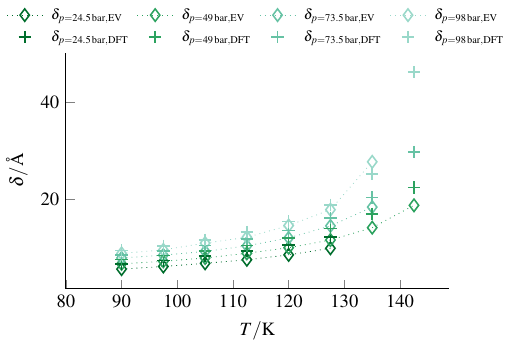}
  \else
    \tikzsetnextfilename{Interface_thickness}
    \input{Interface_thickness}
  \fi

  \vspace{-2em}
  \caption{10-90 interface thickness of the simulations. The dotted lines serve as eye-guides, connecting points with the same pressure. The pluses are the thicknesses of the DFT (\(T,p\)-case) results.}
  \label{fig:Int_thick}
  \vspace{-1em}
\end{figure}

\begin{figure*}
  \centering
  \subfloat[argon]{
  \ifdefined\importfig
    \includegraphics{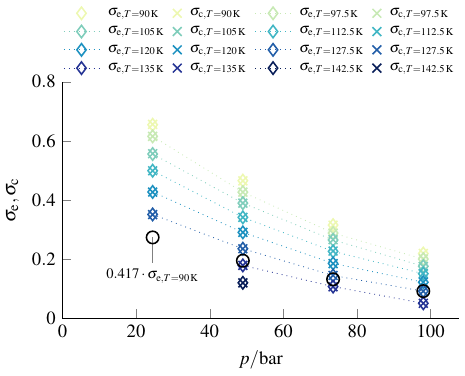}
  \else
    \tikzsetnextfilename{Evap_Ar_p}
    \input{Evap_Ar_p}
  \fi
}
  \subfloat[argon]{
  \ifdefined\importfig
    \includegraphics{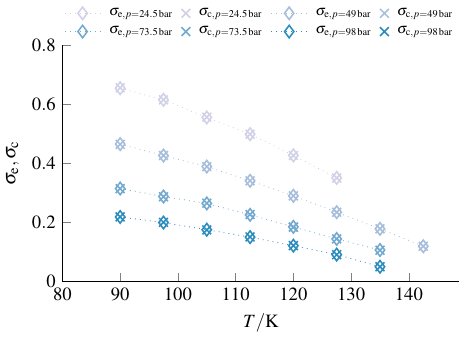}
  \else
    \tikzsetnextfilename{Evap_Ar_T}
    \input{Evap_Ar_T}
  \fi
}\\
  \subfloat[neon]{
  \ifdefined\importfig
    \includegraphics{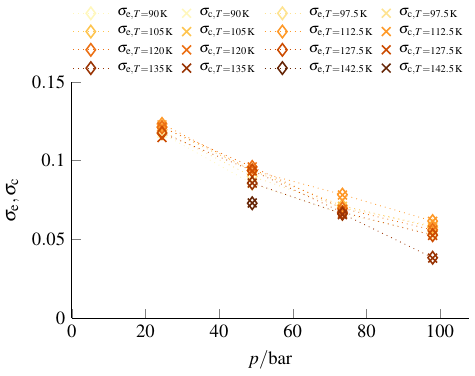}
  \else
    \tikzsetnextfilename{Evap_Ne_p}
    \input{Evap_Ne_p}
  \fi
}
  \subfloat[neon]{
  \ifdefined\importfig
    \includegraphics{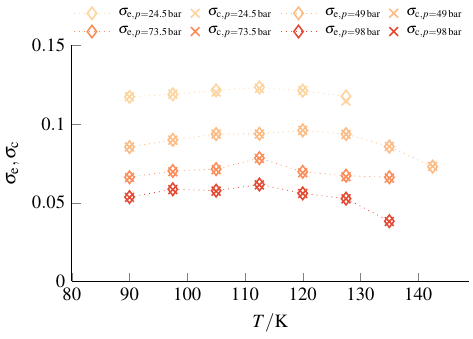}
  \else
    \tikzsetnextfilename{Evap_Ne_T}
    \input{Evap_Ne_T}
  \fi
}
  \caption{Evaporation coefficients of argon and neon at different temperatures and pressures. The dotted lines are eye-guides to connect data with same pressures or temperatures.}
  \label{fig:evap_saft}
\end{figure*}

\autoref{fig:Int_thick} depicts the 10-90 thickness shown in \autoref{eq:Int_fit} of the simulations in comparison with the DFT (\(T,p\)-case) results. Both data sets show the same trends, but the DFT interfaces are approximately \SI{13}{\percent} larger than the EV thicknesses, with a maximum and minimum deviation of \SI{20}{\percent} and \SI{6}{\percent}, respectively. There are three explanations for this discrepancy. First, the models use different interaction potentials. Second, the liquid phase has a slightly different composition. Third, the DFT method does not perfectly align with the MD results \cite{nitzke_2023}. $1 / \delta$ decreases linearly with temperature, which is consistent with the literature \cite{frezzotti_mean_2005}, and $\delta$ rises almost with pressure. This contrasts with the simulations of \citet{ohashi_evaporation_2020}, in which the interface thickness remains nearly constant as the pressure varies. This change in behavior is due to the stronger attraction of neon than the special `neon', and therefore the significant amount of dissolved neon in the liquid. Consequently, the liquid's potential well is weakened, causing the interface to thicken.

\subsection{Evaporation Coefficients of Argon-Neon Mixtures}
\label{ssec:Evap}

The evaporation and condensation coefficients are sampled for the simulations of the argon-neon mixtures. The results are shown in \autoref{fig:evap_saft}.
The evaporation coefficients of argon over pressure are on a lower level than in the simulations with the special `neon', but follow the same exponential trend for all temperatures. The evaporation coefficients over temperature follow mostly a linear trend, in agreement with \cite{ishiyama_molecular_2004}. For low temperatures, the slope decreases slightly. The black circles display $\sigma_\text{e,T=\SI{90}{\kelvin}}$ scaled to match $\sigma_\text{e,T=\SI{127.5}{\kelvin}}$ at \SI{98}{\bar}. The offset of both data series at the other data points shows that the coupled temperature and pressure dependence is not a superposition of both principal dependencies.

\begin{figure*}
  \centering
  \subfloat[\SI{24.5}{\bar} and \SI{90}{\kelvin}   ]{
  \ifdefined\importfig
    \includegraphics{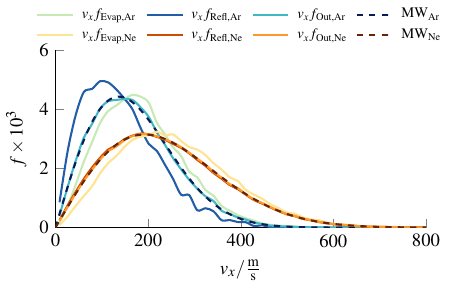}
  \else
    \tikzsetnextfilename{Distr_Func_24.5_090}
    \input{Distr_Func_24.5_090}
  \fi
}
  \subfloat[\SI{24.5}{\bar} and \SI{127.5}{\kelvin}]{
  \ifdefined\importfig
    \includegraphics{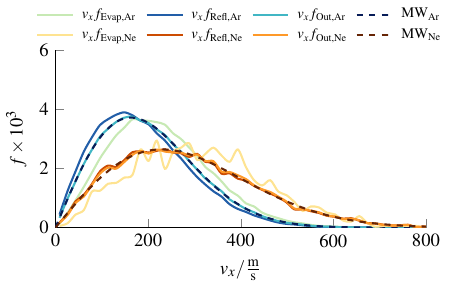}
  \else
    \tikzsetnextfilename{Distr_Func_24.5_127.5}
    \input{Distr_Func_24.5_127.5}
  \fi
}\\
  \subfloat[\SI{49}{\bar} and \SI{90}{\kelvin}   ]{
  \ifdefined\importfig
    \includegraphics{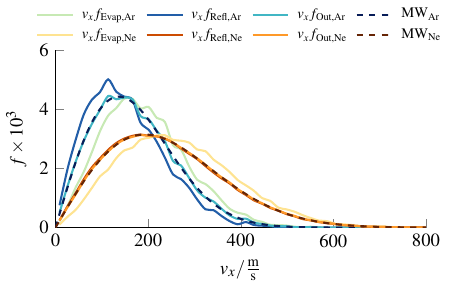}
  \else
    \tikzsetnextfilename{Distr_Func_49_090}
    \input{Distr_Func_49_090}
  \fi
}
  \subfloat[\SI{49}{\bar} and \SI{127.5}{\kelvin}]{
  \ifdefined\importfig
    \includegraphics{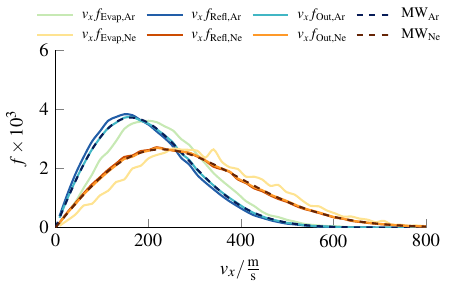}
  \else
    \tikzsetnextfilename{Distr_Func_49_127.5}
    \input{Distr_Func_49_127.5}
  \fi
}\\
  \subfloat[\SI{73.5}{\bar} and \SI{90}{\kelvin}   ]{
  \ifdefined\importfig
    \includegraphics{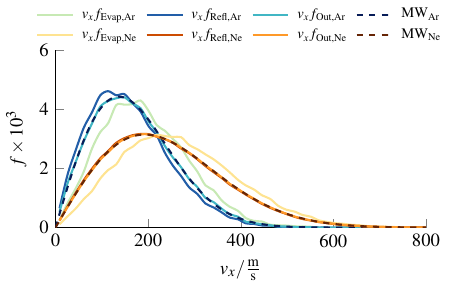}
  \else
    \tikzsetnextfilename{Distr_Func_73.5_090}
    \input{Distr_Func_73.5_090}
  \fi
}
  \subfloat[\SI{73.5}{\bar} and \SI{127.5}{\kelvin}]{
  \ifdefined\importfig
    \includegraphics{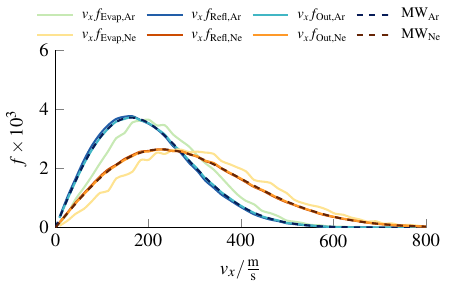}
  \else
    \tikzsetnextfilename{Distr_Func_73.5_127.5}
    \input{Distr_Func_73.5_127.5}
  \fi
}\\
  \subfloat[\SI{98}{\bar} and \SI{90}{\kelvin}   ]{
  \ifdefined\importfig
    \includegraphics{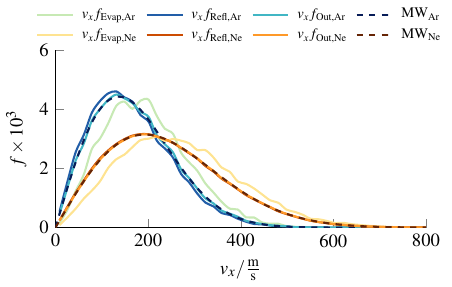}
  \else
    \tikzsetnextfilename{Distr_Func_98_090}
    \input{Distr_Func_98_090}
  \fi
}
  \subfloat[\SI{98}{\bar} and \SI{127.5}{\kelvin}]{
  \ifdefined\importfig
    \includegraphics{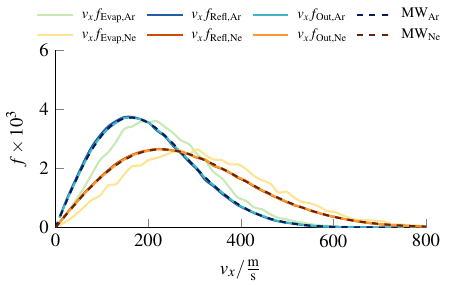}
  \else
    \tikzsetnextfilename{Distr_Func_98_127.5}
    \input{Distr_Func_98_127.5}
  \fi
}
  \caption{Sampled velocity distribution functions of the evaporated, reflected, and outgoing particles of argon and neon at the gaseous boundary. The dashed lines are the Maxwellians of the of the respective temperature of both species.}
  \label{fig:EvapDistrFunc}
\end{figure*}

The evaporation coefficients for neon suggest an exponential decay with respect to pressure. However, this is at a much higher level than in the simulations with the special `neon'. This contrasts with the simulations of \citet{ohashi_evaporation_2020}, in which the special `neon' evaporation coefficients are nearly constant (see \autoref{fig:ohashi_neon_evap}). The authors assume this is due to the increasing interface thickness with increasing pressure. Therefore, increasing pressure also hinders the evaporation process of neon particles. The evaporation coefficients are mostly constant over temperature, with a slight peak around 112.5 K. As the temperature approaches the critical temperature of the mixture, they drop significantly.

The respective distribution functions of evaporated and reflected particles are depicted in \autoref{fig:EvapDistrFunc}. The total outgoing particle flux $v_xf_\text{out}$ is in good agreement with the flux of a Maxwellian distribution with the temperature of the respective simulation. The evaporation and reflected particle distribution functions differ from each other. Both seem to follow a Maxwellian flux trend, too, but with different temperatures, where $f_\text{Evap}$ has a higher temperature, but is not exactly the flux of a Maxwellian. $f_\text{Out}$ can be assembled by superposition of $f_\text{Evap}$ and $f_\text{Refl}$ with the evaporation coefficient. A similar finding has also been reported in EV \cite{ohashi_evaporation_2020} and MD simulations \cite{kon_method_2014}.


\section{Conclusions}
\label{sec:conc}

In this paper, we have introduced and validated a novel multi-species Enskog-Vlasov solver developed specifically to determine evaporation coefficients of fluids in high-pressure environments. This addresses a significant gap in the available modeling of Lamanna \emph{at al.} \cite{Lamanna_2024}, which is necessary for efficient fuel mixing in internal combustion engines.

The innovative aspect of our solver lies in its method of separately handling collisions for different species to identify each species before calculating collisions. This approach is essential for the correct determination of collision displacements, before the second collision particle is selected. We introduced a novel pair correlation function at contact derived from the BMCSL equation of state. However, the compliance of this approach with Onsager relations remains unconfirmed, as it neither strictly violates them nor explicitly incorporates known corrections. This aspect needs further investigation in subsequent studies.

Through comprehensive numerical simulations, we demonstrated the solver's effectiveness across various scenarios. Initial tests showed good agreement with molecular dynamics results during binary fluid relaxation, reinforcing the solver's capability to accurately model fluid dynamics. Further, comparisons against established state-of-the-art results validated the solver’s accuracy in predicting the liquid-vapor equilibrium and evaporation coefficients under varying pressure conditions.

Particularly noteworthy are the outcomes obtained when utilizing realistic particle diameters and masses for argon-neon mixtures. These simulations highlighted substantial differences from conventional single-species methods, especially for the evaporation of neon, which is the trace species in the liquid bulk. This underlines the critical importance of considering species-specific interactions for precise modeling of multi-species evaporation processes.

Additionally, our results closely aligned with SAFT-VRQ Mie and classical density functional theory, providing robust cross-validation and reinforcing the reliability of the proposed solver. The solver successfully predicted liquid and vapor compositions and density profiles of the liquid-vapor interface.

Detailed analyses with the SAFT-VRQ Mie comparison cases further revealed the intricate dependencies of evaporation and condensation coefficients on temperature and pressure. These dependencies of argon were consistent with previous findings, exhibiting a near-linear trend with temperature and an exponential decline with increasing pressure. But the data cannot be fully explained by a superposition of both dependencies.

Overall, this work significantly advances computational modeling capabilities for multi-species fluid evaporation in high-pressure conditions. The presented solver not only provides necessary evaporation coefficients crucial for modeling of evaporation processes but also offers a versatile framework that can be extended to various other multi-phase and multi-component fluid dynamics problems in future research.


\section*{ACKNOWLEDGEMENTS}

The authors gratefully acknowledge the Deutsche Forschungsgemeinschaft (DFG, German Research Fundation) for funding this research within the project `Droplet Interaction Technologies (GRK 2160/2: DROPIT)' and the Center for Digitalization and Technology Research of the Armed Forces of Germany (dtec.bw) through the project Macro/Micro-Simulation of Phase Decomposition in the Transcritical Regime (MaST); dtec.bw is funded by the European Union--NextGenerationEU.

\section*{Author Declarations}
\subsection*{Conflict of Interest}

The authors have no conflicts to disclose.

\subsection*{Author Contributions}
\textbf{Raphael Tietz:} Conceptualization (lead); Data Curation (lead); Formal Analysis (lead); Investigation (lead); Methodology (lead); Software (lead); Validation (lead); Visualization (lead); Writing -- Original Draft (lead); Writing -- Review \& Editing (equal);
\textbf{Rolf Stierle:} Data Curation (supporting); Investigation (supporting); Methodology (supporting); Software (supporting); Validation (supporting); Writing -- Original Draft (supporting); Writing -- Review \& Editing (equal);
\textbf{Kim Sophie Ellenberger:} Writing -- Review \& Editing (equal);
\textbf{Stefanos Fasoulas:} Funding Acquisition (lead); Project Administration (lead); Supervision (equal); Writing -- Review \& Editing (equal);
\textbf{Marcel Pfeiffer:} Project Administration (supporting); Supervision (equal); Writing -- Review \& Editing (equal);

\section*{Data Availability Statement}

The data that support the findings of this study are available from the corresponding author upon reasonable request.

\bibliography{references}

\end{document}